\documentclass[a4paper,10pt]{article}
\usepackage{amsmath,amsfonts}
\usepackage{times}
\usepackage[T1]{fontenc}
\usepackage{epsfig}
\usepackage{amssymb}
\usepackage[english]{babel}

\usepackage[usenames,dvipsnames]{color}
\usepackage{color}

\begin{document}

\title{\Huge{\bf A BPS Skyrme model}
}

\author{C. Adam$^{a)}$\thanks{adam@fpaxp1.usc.es},
J. S\'{a}nchez-Guill\'{e}n
$^{a)}$\thanks{joaquin@fpaxp1.usc.es}, A. Wereszczy\'{n}ski
$^{b)}$\thanks{wereszczynski@th.if.uj.edu.pl}
\\ \\ 
$^{a)}$ Departamento de F\'isica de Part\'iculas, Universidad
\\ 
de Santiago, and Instituto Galego de F\'isica de Altas Enerx\'ias
\\ 
(IGFAE) E-15782 Santiago de Compostela, Spain
\\ \\
$^{b)}$ Institute of Physics,  Jagiellonian University,
\\ 
Reymonta 4, Krak\'{o}w, Poland}

\maketitle

\begin{abstract}
Within the set of generalized Skyrme
models, we identify a submodel which has both infinitely many symmetries and a Bogomolny bound which is saturated by infinitely many exact soliton solutions. Concretely, the submodel
consists of the square of the baryon
current and a potential term only.  Further, already on the
classical level, this BPS Skyrme model reproduces some features of the 
liquid drop model
of nuclei. Here, we review the properties of the model and we discuss
 the semiclassical quantization of the simplest Skyrmion (the nucleon).
\\ \\ \\ \\ \\
{\large Contribution to the Conference Proceedings of the 28th International Colloquium On
Group
  Theoretical Methods In Physics (GROUP 28), July 2010, Northumbria,
England.
}

\end{abstract}

\newpage
\section{Introduction}
The derivation of the correct low energy hadron dynamics from the underlying 
fundamental theory  (QCD)
is undoubtedly one of the most prominent challenges of strong interaction physics, due to the non-perturbative nature of the quark and gluon
interactions in the low energy limit. In the large $N_c$ limit
it is known that QCD becomes equivalent to an effective theory of mesons 
\cite{thooft}, where Baryons (hadrons as well as atomic nuclei) appear as 
solitonic excitations with an identification between the baryon number 
and the topological charge \cite{witten1}.  One of the most popular 
realizations of this idea is the Skyrme model \cite{skyrme} i.e., a 
version of a phenomenological chiral Lagrangian, where the primary 
ingredients are meson fields. 
The original 
Lagrangian has the following form
\begin{equation}
L=L_2 + L_4 + L_0,
\end{equation}
where to the sigma model part
\begin{equation}
L_2=-\frac{f_{\pi}^2}{4} \; \mbox{Tr} \; (U^{\dagger} \partial_{\mu} U \; 
U^{\dagger}
\partial^{\mu} U  )
\end{equation}
a quartic term, referred as Skyrme term, has to be added, to
circumvent the standard Derrick argument for the non-existence of
static solutions
\begin{equation}
L_4= - \frac{1}{32 e^2}\; \mbox{Tr} \; ([U^{\dagger} \partial_{\mu}
U,U^{\dagger} \partial_{\nu} U]^2 ).
\end{equation}
The last term is a potential
\begin{equation}
L_0 = -\mu^2 V(U,U^{\dagger}),
\end{equation}
which explicitly breaks the the chiral symmetry. Its particular form is 
usually adjusted to a concrete physical
situation. Finally, static properties of baryons as well as nuclei are derived with the help of 
the semiclassical quantization of the solitonic zero modes \cite{nappi1}, 
\cite{nappi2}, \cite{sutcliffe1}. 
\\
Within this framework, it has been established that topological solitons (Skyrmions) possess 
qualitative properties of baryons \cite{nappi1}, \cite{nappi2}. Quantitatively, for the original Skyrme model nucleon properties are reproduced within a typical precision of about 30\%.
For higher nuclei, one of the principal problems of the Skyrme model is that the binding energies of the nuclei result too large already at the classical level. This problem may be traced back to the fact that the solitons in the Skyrme model are not BPS, i.e., they do not saturate the Bogomolny bound of the model. Other related problems are too strong internuclear forces at intermediate distances and the cristal like behaviour of large nuclei in the standard Skyrme model.
The situation seems to be better for the excitation spectra of nuclei. Indeed, by identifying the symmetries of higher Skyrmions and by the use of the semiclassical quantization method the (iso-) rotational excitations of nuclei have been described rather successfully in  \cite{sutcliffe1}.      
\\
In any case, if interpreted as an effective field theory, the Skyrme model should be viewed as a
derivative expansion, where higher derivative terms have
been neglected.  Further, baryons, i.e., extended (solitonic) 
solutions, have regions where derivatives are not small, so it may not be justified to
omit such terms, and one should take them into account.
Indeed, many {\it generalized} Skyrme models, i.e., extensions 
of the original Skyrme  Lagrangian to models containing higher derivative 
terms have been investigated \cite{adkins}-\cite{vec skyrme}.
\\
Among them, the simplest and most natural generalization of the Skyrme model 
is defined by the addition of
the following sextic derivative term
\begin{equation}
L_6=\frac{\lambda^2}{24^2} \;\left[ \mbox{Tr} \; (\epsilon^{\mu
\nu \rho \sigma} U^{\dagger} \partial_{\mu} U \; U^{\dagger}
\partial_{\nu} U \; U^{\dagger} \partial_{\rho} U) \right]^2. \label{6}
\end{equation}
Phenomenologically, this term effectively appears if one considers a massive 
vector field coupled to the chiral field via the baryon density 
\cite{adkins}, \cite{vec skyrme}.  It turns out that such a modified 
Skyrme model gives slightly better quantitative predictions \cite{adkins}, 
\cite{jackson}, \cite{piette1}, \cite{ding}. Besides, this sextic term is at most quadratic in time derivatives, so it allows for a standard hamiltonian interpretation.  It is, in fact, nothing but
the topological (or baryon) current density squared,
\begin{equation}
L_6=\lambda^2 \pi^4 \mathbb{B}_\mu \mathbb{B}^\mu    
\end{equation}
where
\begin{equation}
 \mathbb{B}^\mu = \frac{1}{24\pi^2}   \mbox{Tr} \; (\epsilon^{\mu
\nu \rho \sigma} U^{\dagger} \partial_{\nu} U \; U^{\dagger}
\partial_{\rho} U \; U^{\dagger} \partial_{\sigma} U)   .
\end{equation}
\\
Within the set of generalized Skyrme models, the standard Skyrme model may be viewed as an approximation which is mode tractable and may be sufficient for the description of some physical properties of nuclei. Here we propose another approximation which consists of the potential and the sextic term only,  which we will call 
{\it the BPS Skyrme model} 
 \begin{equation}
L_{06}=\frac{\lambda^2}{24^2 } \;\left[  \mbox{Tr} \;  ( \epsilon^{\mu 
\nu \rho \sigma} U^{\dagger} \partial_{\mu} U \;
U^{\dagger} \partial_{\nu} U \;
U^{\dagger} \partial_{\rho} U) \right]^2 - \mu^2 V(U,U^{\dagger}).
\end{equation}
The subindex 06 refers to the fact that in the above Lagrangian only a
potential term without derivatives and a term sextic in 
derivatives are present. 
This model 
is by construction more topological 
in nature than any of the standard versions of the Skyrme model. 
It shares with the standard Skyrme model the stabilization of soliton
solutions by a higher order term in derivatives (a sextic term in our case).
It is an 
example of a BPS theory with topological solitons saturating the pertinent 
Bogomolny bound and therefore, with zero binding energies at the classical level. 
Further, this 
BPS model possesses a huge number of symmetries which lead to its 
integrability. What is more important, among its symmetries are the volume 
preserving diffeomorphisms on base space. This allows to interpret the 
BPS Skyrme matter as an incompressible liquid. Moreover, the solitons of this 
theory are of compact type which results (classically) in a finite range of interactions 
(contact-like interactions).   
As a consequence, the BPS Skyrme model apparently is a good guess 
for certain properties of nuclei.
\\
We want to remark that there exists another BPS generalization 
of the Skyrme model recently proposed by Sutcliffe \cite{sutcliffe bps} based 
on dimensional reduction of the (4+1) dimensional 
Yang-Mills (YM) theory, where the $SU(2)$ 
Skyrme field is accompanied by an infinite tower of vector and tensor 
mesons. 
\section{The BPS Skyrme model}
\subsection{Field equations}
We use the standard parametrization of $U$ by means of a real scalar $\xi$ 
and a three component unit vector $\vec{n}$ field 
($\vec \tau$ are the Pauli matrices), 
$$
U=e^{i \xi \vec{n} \cdot \vec{\tau}}.
$$
The vector field may be related to a complex scalar $u$ by the 
stereographic projection
$$
\vec{n}=\frac{1}{1+|u|^2} \left( u+\bar{u}, -i ( u-\bar{u}),
|u|^2-1 \right).
$$
For the topological current the resulting expression is
$$
\mathbb{B}^\mu = \frac{-i}{\pi^2}\frac{  \sin^4 \xi}{(1+|u|^2)^4} \; 
\epsilon^{\mu \nu \rho \sigma} \xi_{\nu} u_{\rho} \bar{u}_{\sigma} ,
$$
and, assuming for the potential
$
V=V({\rm tr} (U+U^\dagger) )=V(\xi)
$
(which we assume for the rest of the paper) we get  for the lagrangian  
\begin{equation}
L_{06}= -\frac{  \lambda^2 \sin^4 \xi}{(1+|u|^2)^4} \;\left(  
\epsilon^{\mu \nu \rho \sigma} \xi_{\nu} u_{\rho} \bar{u}_{\sigma} \right)^2
-\mu^2 V(\xi) .
\end{equation}
The Euler--Lagrange equations read
$$ \frac{\lambda^2 \sin^2 \xi}{(1+|u|^2)^4} \partial_{\mu} ( \sin^2 \xi \; 
H^{\mu}) + \mu^2 V'_{\xi}=0,$$
$$ \partial_{\mu} \left( \frac{K^{\mu}}{(1+|u|^2)^2} \right)=0,$$
where $$ H_{\mu} = \frac{\partial (  \epsilon^{\alpha \nu \rho \sigma} 
\xi_{\nu} u_{\rho} \bar{u}_{\sigma})^2}{ \partial \xi^{\mu}}, \;\;\; K_{\mu} = 
\frac{\partial (  \epsilon^{\alpha \nu \rho \sigma} \xi_{\nu} u_{\rho} \bar{u}
_{\sigma})^2}{\partial \bar{u}^{\mu}}.$$
These objects obey the useful formulas 
$$H_{\mu} u^{\mu}=H_{\mu} \bar{u}^{\mu}=0, 
\; K_{\mu}\xi^{\mu}=K_{\mu} u^{\mu}=0, \;\; H_{\mu} \xi^{\mu}=K_{\mu} 
\bar{u}^{\mu} = 2 (  \epsilon^{\alpha \nu \rho \sigma} \xi_{\nu} u_{\rho} 
\bar{u}_{\sigma})^2. $$ 
\subsection{Symmetries}
Apart from the standard Poincare symmetries, the model has an infinite 
number of
target space symmetries. The sextic term alone is the square of the pullback 
of the volume form on the target space $S^3$, where this target space volume
form reads explicitly
\begin{equation}
dV= -i\frac{\sin^2 \xi}{(1+|u|^2)^2}d\xi du d\bar u
\end{equation}
and the exterior (wedge) product of the differentials is understood.  
Therefore, the sextic term alone is invariant under all target space
diffeomorphisms which leave this volume form invariant (the volume-preserving
diffeomorphisms on the target $S^3$). The potential term in general does not
respect all these symmetries, but depending on the specific choice, it may
respect a certain subgroup. Concretely, for
$V=V(\xi)$, the potential is invariant under those volume-preserving 
target space diffeomorphisms which do not change $\xi$, 
\begin{equation}
\xi \to \xi \, , \quad u\to \tilde u(u,\bar u, \xi) \, ,\quad 
(1+|\tilde u|^2)^{-2} d\xi d\tilde u d\bar{\tilde u} = 
(1+| u|^2)^{-2} d\xi  u d\bar{ u} .
\nonumber
\end{equation}
Since $u$ spans a two-sphere in target space,
these transformations form a one-parameter family of the groups of the 
area-preserving
diffeomorphisms on the corresponding target space $S^2$ (one-parameter family
because the transformations may still depend on $\xi$, although they act
nontrivially only on $u, \bar u$). Both the Poincare
transformations and this family of
area-preserving target space diffeomorphisms are
symmetries of the full action, so they are Noether symmetries with the
corresponding conserved currents. The latter symmetries may, in fact, be
expressed in terms of the generalized integrability, as we briefly discuss in
the next subsection. 
\\
The energy functional for static fields has an additional group of infinitely
many symmetry transformations, as we want to discuss now. These symmetries are
not symmetries of the full action, so they are not of the Noether type, but
nevertheless they are very interesting from a physical point of view, as we
will see below. The energy functional for static fields reads
\begin{equation}
E=\int d^3 x \left(  \frac{\lambda^2 \sin^4 \xi}{(1+|u|^2)^4} 
(\epsilon^{mnl} i \xi_m u_n\bar{u}_l)^2 +\mu^2 V \right)
\end{equation}
and we observe that both $d^3 x$ and $\epsilon^{mnl} i \xi_m u_n\bar{u}_l$ are
invariant under coordinate transformations of the base space coordinates $x_j$
which leave the volume form $d^3 x$ invariant. So this energy functional has
the volume-preserving diffeomorphisms on base space as symmetries. These
symmetries are precisely the symmetries of an incompressible ideal fluid,
which makes them especially interesting in the context of applications to
nuclear matter. Indeed, the resulting field theory is able to reproduce some
basic features of the liquid droplet model of nuclei, see
e.g. \cite{proposal}, \cite{slobo} and the discussion below. 
\subsection{Integrability}
The BPS Skyrme model is integrable in the sense that there 
are infinitely 
many conserved charges. Indeed, it belongs to a family of models integrable 
in the sense of the generalized integrability \cite{alvarez},
\cite{pullback}. 
To show that we introduce 
$$ \mathcal{K}^{\mu}=\frac{K^{\mu}}{(1+|u|^2)^2}.$$
The currents are
$$ J_{\mu}=\frac{\delta G}{\delta \bar{u}} \bar{\mathcal{K}}^{\mu} - 
\frac{\delta G}{\delta u} \mathcal{K}^{\mu}, \;\;\; G=G(u,\bar{u},\xi) $$
where $G(u,\bar u,\xi)$ is an arbitrary real function of its arguments.
Then, 
$$\partial^{\mu} J_{\mu}= G_{\bar{u}\bar{u}} \bar{u}_{\mu} 
\bar{\mathcal{K}}^{\mu} + G_{\bar{u}u} u_{\mu} \bar{\mathcal{K}}^{\mu} + 
G_{\bar{u}} \partial_{\mu} \bar{\mathcal{K}}^{\mu} -G_{u \bar{u}}
\bar{u}_{\mu} 
\mathcal{K}^{\mu} -G_{uu} u_{\mu} \mathcal{K}^{\mu} - G_{u} \partial_{\mu} 
\mathcal{K}^{\mu}$$ $$ + G_{\bar{u} \xi} \xi_{\mu} \bar{\mathcal{K}}^{\mu} - 
G_{u \xi} \xi_{\mu} \mathcal{K}^{\mu}=0, $$
where we used that $u_{\mu} \mathcal{K}^{\mu}=\xi_{\mu} \mathcal{K}^{\mu}=
0 $, $ \bar{u}_{\mu} \mathcal{K}^{\mu}=u_{\mu} \bar{\mathcal{K}}^{\mu},$ which 
follow from the previous identities. Finally using the field equations for 
the complex field, 
one, indeed, finds an infinite number of conserved currents.
These currents are a higher dimensional generalization of 
those constructed for the pure baby Skyrme model \cite{integr} and are 
generated by the relevant subgroup of the 
volume-preserving diffeomorphisms on the target space as discussed in the
previous subsection, see also
\cite{ab-dif}. 
\subsection{Bogomolny bound}
In the BPS Skyrme model, there exists a Bogomolny bound for static finite energy solutions
\cite{}. The most concise version of the proof has been given by M. Speight \cite{}, and here we use this version of the proof. In a first step, the energy functional has to be rewritten as a complete square plus a remainder,
\begin{eqnarray}
E &=& \int d^3 x \left(  \frac{\lambda^2 \sin^4 \xi}{(1+|u|^2)^4} 
(\epsilon^{mnl} i \xi_m u_n\bar{u}_l)^2 +\mu^2 V \right) \nonumber \\
&=& \int d^3 x \left( \frac{\lambda \sin^2 \xi}{(1+|u|^2)^2} 
\epsilon^{mnl} i \xi_mu_n\bar{u}_l 
\pm  \mu \sqrt{V} \right)^2 \nonumber \\
&\mp & \int d^3 x 
\frac{2\mu \lambda \sin^2 \xi \sqrt{V}}{(1+|u|^2)^2} \epsilon^{mnl} i \xi_m 
u_n \bar{u}_l \nonumber \\
&\geq &  \pm (2\lambda \mu \pi^2 )\left[ \frac{-i}{\pi^2}
\int d^3 x \frac{ \sin^2 \xi \sqrt{V}}{(1+|u|^2)^2} 
\epsilon^{mnl}  \xi_m u_n \bar{u}_l \right]  \equiv E_{\rm BPS} ,
\end{eqnarray}
and in a second step it remains to prove that the remainder $E_{\rm BPS}$ is topological and does  not depend on the field configuration. For this we note that the integrand in the above expression is not only a three-form on the base space but, in addition, it is the pullback under the map $U$ of a three-form on target space. Therefore, the base space integral of the base space three-form is equal to the target space integral of the target space three-form times the number of times the target space is covered while the base space is covered once (i.e., times the winding number $|B|$), 
\begin{eqnarray}
E_{\rm BPS} &=& 2\lambda \mu \pi^2 |B| \frac{-i}{\pi^2} \int_{S^3} \frac{ d\xi du d\bar u \sin^2 \xi }{(1+|u|^2)^2} \sqrt{V(\xi)} \nonumber \\
& \equiv & 2\lambda \mu \pi^2 <\sqrt{V}>_{S^3} |B|  
\end{eqnarray}
where $<\sqrt{V}>_{S^3}$ is the average value of $\sqrt{V}$ on the target space $S^3$,
$$
<\sqrt{V}>_{S^3} = \frac{\int_{S^3} d\Omega \sqrt{V}}{\int_{S^3} d\Omega}
$$
and $d\Omega$ is the volume form on $S^3$ (and $\int_{S^3} d\Omega = \pi^2$).
The corresponding Bogomolny (first order) equation is
\begin{equation}
\frac{\lambda \sin^2 \xi}{(1+|u|^2)^2} \epsilon^{mnl} i \xi_mu_n\bar{u}_l 
= \mp \mu \sqrt{V} 
\end{equation}
and is satisfied by all soliton solutions which we will encounter below.
\subsection{Exact solutions}
We are interested in static topologically nontrivial solutions. Thus $u$ must 
cover the whole complex plane ($\vec{n}$ covers at least once $S^2$) 
and $\xi \in [0,\pi]$. The natural (hedgehog) ansatz is
\begin{equation}
\xi = \xi (r), \;\;\; u(\theta, \phi) = g (\theta) e^{in \phi}.
\end{equation}
Then, the field equation for $u$ reads
$$ 
\frac{1}{\sin \theta} \partial_{\theta} \left( \frac{ g^2g_\theta}{(1+g^2)^2 
\sin \theta} \right) - \frac{gg_\theta^2}{(1+g^2)^2\sin^2 \theta}=0,
$$
and the solution with the right boundary condition is
\begin{equation} \label{u-sol}
g(\theta) = \tan \frac{\theta}{2}.
\end{equation}
Observe that this solution holds for all values of $n$. 
The equation for the real scalar field is 
$$
\frac{n^2\lambda^2 \sin^2 \xi }{2r^2} \partial_r \left(\frac{\sin^2 \xi \; 
\xi_r}{r^2} \right) - \mu^2 V_{\xi}=0.
$$ 
This equation can be simplified by introducing the new variable 
\begin{equation}
z=\frac{\sqrt{2}\mu r^3}{3 |n|\lambda} .
\end{equation} 
It reads
\begin{equation} \label{xi-eq}
\sin^2 \xi \; \partial_z \left(\sin^2 \xi \; \xi_z\right) -  V_{\xi}=0,
\end{equation}
and may be integrated to 
\begin{equation} 
 \frac{1}{2} \sin^4 \xi \; \xi^2_z=V, \label{bps eq}
\end{equation}
where we chose a vanishing integration constant to get finite energy solutions. 
We remark that this first integration of the field equation is just the dimensionally
reduced Bogomolny equation. 
\\
For a further evaluation we have to specify a potential. Here we shall restrict ourselves to the case of the standard Skyrme potential 
\begin{equation}
V=\frac{1}{2}\mbox{Tr} (1-U) \;\; \rightarrow \;\; V(\xi)=1- \cos \xi. 
\end{equation}
Imposing the boundary conditions for
topologically non-trivial solutions we get 
\begin{equation}
\xi = \left\{
\begin{array}{lc}
2 \arccos \sqrt[3]{ \frac{3z}{4} } & z \in \left[0,\frac{4}{3} \right] \\
0 & z \geq \frac{4}{3}.
\end{array} \right. \label{xi sol}
\end{equation}
The solution is of 
the compacton type, i.e., it has a finite support \cite{comp}
(compact solutions of a similar
type in different versions of the
baby Skyrme models have been found in \cite{GP},
\cite{comp-bS}).
The corresponding energy is
\begin{eqnarray}
E &=&  8 \sqrt{2}\pi \mu \lambda |n| \int_0^{4/3} 
\left(1-\left( \frac{3z}{4} \right)^\frac{2}{3} \right)dz = 
\frac{64\sqrt{2} \pi}{15} \mu \lambda |n|  \label{E-skyrmepot}
\end{eqnarray}
and is linear in the topological charge $|B|=|n|$.
The energy density and topological charge density per unit volume are
\begin{eqnarray}
{\cal E}&=&   8 \sqrt{2} \mu \lambda (1- |n|^{-\frac{2}{3}} \tilde r^2 ) 
\quad \mbox{for} \quad 0\le \tilde r \le |n|^\frac{1}{3} \nonumber \\
&=& 0\quad \mbox{for} \quad \tilde r > |n|^\frac{1}{3}. 
\end{eqnarray}
\begin{eqnarray}
{\cal B} &=&  \mbox{sign} (n) \frac{4}{\pi^2} (1- |n|^{-\frac{2}{3}}\tilde
r^2)^\frac{1}{2} \quad \mbox{for} \quad 0\le \tilde r \le 
|n|^\frac{1}{3} \nonumber \\
&=& 0\quad \mbox{for} \quad \tilde r > |n|^\frac{1}{3}
\end{eqnarray}
where
\begin{equation}
\tilde r = \left( \frac{\sqrt{2}\mu}{4  \lambda} \right)^\frac{1}{3} r 
\equiv \frac{r}{R_0} =
\left(\frac{3|n|z}{4}\right)^\frac{1}{3}
\end{equation}
(here $R_0$ is the compacton radius for the soliton with charge one), see Figure 1.
\\
\begin{figure}[h!]
\includegraphics[angle=0,width=0.55 \textwidth]{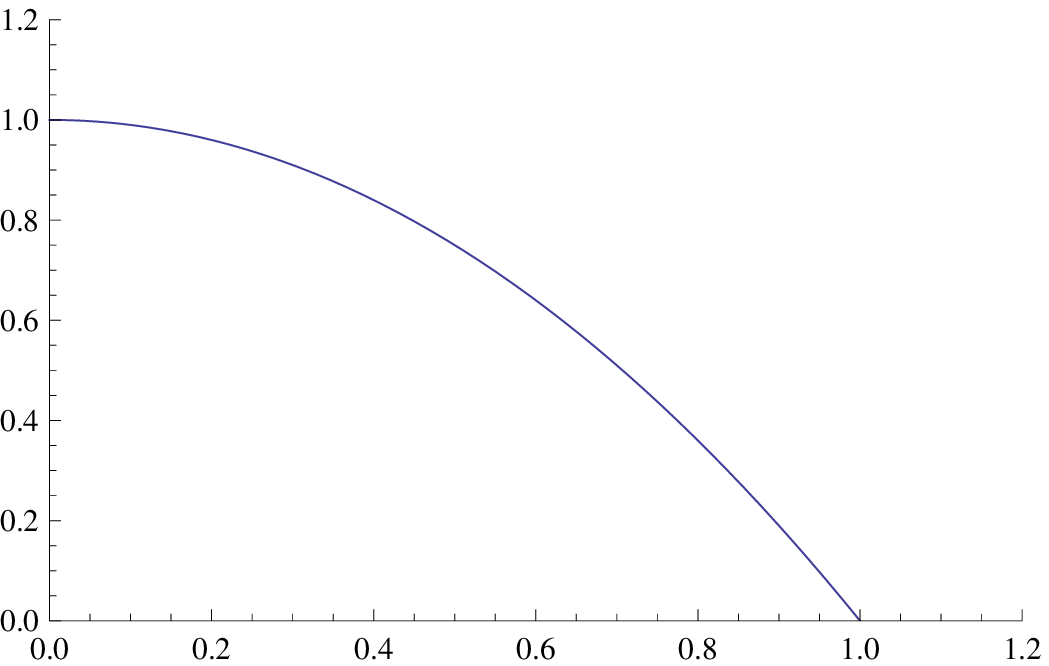}
\includegraphics[angle=0,width=0.55 \textwidth]{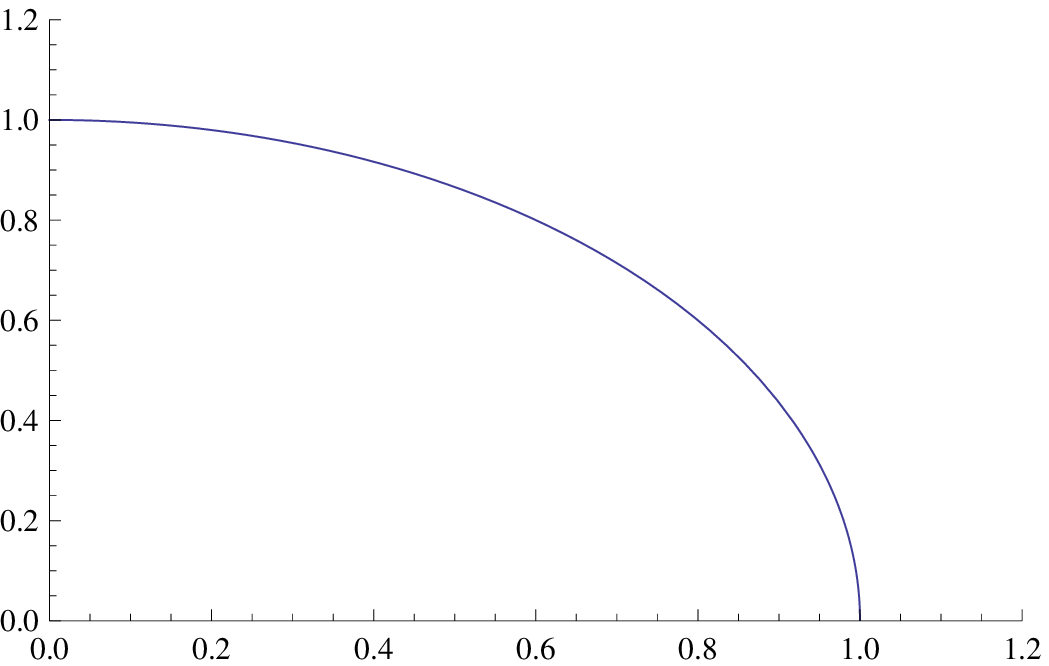}
\caption{Normalized energy density (left figure) and topological charge 
density (right figure) as a function of the
rescaled radius $\tilde r$, for topological charge n=1. For $|n|>1$, the
height of the densities remains the same, whereas their radius grows like
$|n|^\frac{1}{3}$  }\label{rys1}
\end{figure}
\section{Some phenomenology of nuclei}
For simplicity, we choose the standard Skyrme potential $V=1-\cos \xi$ 
throughout this section.
\subsection{Classical aspects}
Already the classical solutions of
the BPS Skyrme model describe quite well some static 
properties of nuclei \cite{proposal}. Concretely, we find
\begin{itemize}
\item A linear mass - baryon number relation: 
 $E=64\sqrt{2} \mu\lambda |B|/15 \equiv E_0 |B|$, 
which is a
well-known relation for nuclei
\item No binding energies due to the
BPS property of soliton solutions.
Binding energies for physical nuclei are small ($\leq 1\%$), while 
in the standard Skyrme model they are significantly bigger.
\item No forces between solitons:
due to the BPS  and compact nature of solutions there are no long range forces.
Physical nuclei have a very short range interaction.
\item Radii of nuclei: the
compact solutions have a definite radius $R = (2\sqrt{2}\lambda |B|/\mu)^{(1/3)}$ $ \equiv
R_0 \sqrt[3]{|B|}$ which, again, is a
well-known  relation for physical nuclei. 
\item Incompressible fluid:
the static energy functional has the volume-preserving diffeos (VPDs) as symmetries, which
are the symmetries of an incompressible ideal fluid. 
Physical nuclei do not have this symmetry: they have definite shape, and deformations cost energy.
But: volume-preserving deformations cost much less energy, as a consequence of the liquid drop model of nuclei.
\end{itemize}
 To summarize, the classical model already reproduces some features of the nuclear liquid drop model (mass and volume proportional to baryon charge, strictly finite size, VPDs). 
But clearly this picture is too naive (there are no pions and, therefore,  no long range interactions and no pion cloud; there are no quantum corrections, etc.). Further, 
these classical results are
probably more trustworthy for not too small nuclei, because 
\begin{itemize}
\item[i)] the contribution of the pion cloud (which is absent in our model) 
to the size of the nucleus is of lesser significance for larger nuclei. We
remind that in addition to the core of a nucleus (with a size which grows
essentially with the third root of the baryon number) a surface term is known
to exist for physical nuclei
whose thickness is essentially independent of the baryon number.
\item[ii)] the description of a nucleus as a liquid drop of nuclear matter is
  more appropriate for larger baryon number.
\item[iii)] the contribution of (iso-) rotational quantum 
excitations to the total mass of
a nucleus is smaller for larger nuclei, essentially because of the larger
moments of inertia of larger nuclei.   
\end{itemize}
\subsection{Quantization}
Here we perform the first steps towards a semiclassical (collective coordinates) quantization of a soliton, as is done for the standard Skyrme model. Concretely, the standard procedure consists in quantizing the (iso-)rotational degrees of freedom of each soliton. In principle, one can quantize all solitons in this way and obtain information about the binding energies and rotational spectra of nuclei, but here we shall restrict to the simplest $B=1$ case (the nucleon).  
Following the standard treatment, we introduce the collective 
coordinates of the isospin by including a time-dependent iso-rotation of the 
classical soliton configuration
\begin{equation}
U(x)=A(t)U_0(x)A^{\dagger}(t),
\end{equation}
where  $A(t)=a_0+ia_i\tau_i \in SU(2)$ and $a_0^2+\vec{a}^2=1$.
We remark that the $B=1$  hedgehog is invariant under a combined rotation in base and
isotopic space, therefore, it is enough to introduce the collective
coordinates of one of the two. 
The standard procedure consists in inserting the iso-rotation into the lagrangian, introducing the conjugate momenta to the SU(2) variables $a_n$ and applying the standard canonical quantization to the latter. This results in an energy expression which is the sum of the static soliton energy $E_0$ and an isospin (or equivalently spin) contribution
$$
H = E_0
+ \frac{\hbar^2 I^2}{2{\cal I}}=E_0+\frac{\hbar^2 S^2}{2{\cal I}},
$$
where $I^2$ is the isospin operator (the spherical laplacian on $S^3$).
Further, ${\cal I}$ is the moment of inertia
\begin{equation}
{\cal I} = \frac{4\pi}{3} \int dr \sin^4 \xi \, \xi_r^2 =
\frac{2^8\sqrt{2}\pi}{15\cdot 7}\lambda \mu \left(\frac{\lambda}{\mu} \right)^\frac{2}{3}
\nonumber
\end{equation}
The soliton with baryon number one is quantized as a fermion. Concretely, the
nucleon has spin and isospin 1/2, whereas the $\Delta$ resonance has spin and
isospin 3/2, so we find for their masses 
\begin{equation}
M_N=E_0+\frac{3\hbar^2}{8{\cal I}}, \;\; M_\Delta=E_0+
\frac{15\hbar^2}{8{\cal I}} \;\; 
\Rightarrow \;\; 
M_\Delta-M_N=\frac{3\hbar^2}{2{\cal I}},
\end{equation}
like in the 
standard Skyrme model. These expressions may now be fitted to the physical masses
of the nucleon ($M_N = 938.9$ MeV) and the $\Delta$ resonance ($M_\Delta =
1232$ MeV), which determines the coupling
constants
$$
\lambda \mu = 45.70 {\rm MeV} \, , \qquad 
\frac{\lambda}{\mu} = 0.2556 {\rm fm}^3 
$$
These may now be used to ``predict''
further physical quantities like, e.g. the charge radii of the nucleons.
For the isoscalar (baryon) charge density per unit $r$ we find
\begin{equation}
\rho_0 = 4\pi r^2 B^0= - \frac{2}{\pi} \sin^2 \xi \xi'_r 
\end{equation}
and for the isovector charge density per unit $r$
\begin{equation}
\rho_{1}=\frac{1}{{\cal I}} \frac{4\pi}{3} \lambda^2  \sin^4\xi \xi'^2_r.
\end{equation}
The corresponding isoscalar and isovector mean square electric radii are
\begin{equation}
<r^2>_{e,0}= \int dr r^2 \rho_0 = \left( \frac{ \lambda}{\mu} \right)^{2/3},
\end{equation}
\begin{equation}
<r^2>_{e,1}= \int dr r^2 \rho_1 = 
\frac{10}{9} \left( \frac{ \lambda}{\mu} \right)^{2/3}.
\end{equation}
Further, the isoscalar magnetic radius is defined as the ratio
\begin{equation}
< r^2>_{m,0} = \frac{\int dr r^4 \rho_0}{\int dr r^2 \rho_0} = \frac{5}{4}
\left( \frac{ \lambda}{\mu} \right)^{2/3}.
\end{equation}
The numerical values are displayed in Table \ref{table2}.  
\\
\begin{table}
\begin{center}
\begin{tabular}{|c|c|c|c|}
\hline
radius & BPS & massive  & exp.  \\
 & Skyrme & Skyrme & \\
\hline 
compacton &  0.90 & -  & - \\
electric isoscalar $r_{e,0}$ &  0.64 &  0.68 & 0.72 \\
electric isovector $r_{e,1}$ &  0.67  &  1.04 & 0.88 \\
magnetic isoscalar $r_{m,0}$ &  0.71 &  0.95 & 0.81 \\
\hline
$r_{e,1}/r_{e,0}$ & 1.05 & 1.53 & 1.22 \\
$r_{m,0}/r_{e,0}$ & 1.12 & 1.40 & 1.13 \\
$r_{e,1}/ r_{m,0}$ & 0.94 & 1.10 & 1.09 \\
\hline
\end{tabular}  
\caption{Compacton radius and some charge radii and their ratios
for the nucleon. The numbers
  for the massive Skyrme model  are from Ref. \cite{nappi2}.
All radii are in fm.} \label{table2}
\end{center}
\end{table}
Firstly, in the BPS Skyrme model all radii are
bound by the compacton radius $R_0= \sqrt{2}(\lambda /\mu)^{(1/3)}$. 
Secondly, all radii in the BPS Skyrme model
are significantly
smaller than their physical values, as well as significantly smaller than the
values predicted in the standard massive Skyrme model. This has
to be expected, because we know already that the pion cloud is absent in the
BPS Skyrme model. 
Thirdly, we see that the error due to the 
 absence of the pion cloud in the model 
 partly cancels in the ratios, as one would expect. 
\section{Conclusions}
In this work we proposed a submodel within the set of 
generalized Skyrme models which 
we called {\it the BPS Skyrme model}. The model is quite interesting on its own,  because of its infinitely many symmetries and conserved charges
charges, its solvability for any form of the potential, and because all
solutions are of the BPS type. They obey a first order differential 
equation and saturate a Bogomolny bound. 
\\
Further, we found some evidence that the model may be a good starting point for an effective field theory description of nuclei. For a further development of tis application of the model, however, the following problems have to be resolved or investigated.
 \begin{itemize}
 \item Symmetry breaking: the infinitely many symmetries of the model are not shared by physical nuclei. In addition, it is not clear how to select the correct soliton from the infinitely many ones related by symmetries or how to quantize these infinitely many symmetries. 
Therefore, a realistic phenomenological application will require the breaking of these symmetries. The challenge will be to identify a breaking mechanism which effectively breaks the unwanted symmetries without perturbing too much the good properties of the model (like weak binding energies, weak internuclear forces).
\item Quantization of higher nuclei: the semiclassical quantization of higher solitons should be performed and applied to higher nuclei, along the lines of what was done for the standard Skyrme model, e.g., in \cite{sutcliffe1}. Recently in  \cite{Bon-Mar},
the authors used a version of the BPS Skyrme model with a different potential 
for this purpose.
Concretely, they calculated the exact static soliton solutions plus the (iso-) rotational energies in the rigid rotor quantization for general baryon charge $B=n$ for the spherically symmetric ansatz. Then they allowed for small contributions to the total energies from the quadratic and quartic Skyrme terms and fitted the resulting binding energies to the experimental binding energies of the most abundant isotopes of higher nuclei, assuming, as is usually done, that these correspond to the states with the lowest possible value of the isospin. The resulting  agreement between calculated and experimentally determined masses and binding energies is impressive, lending further support to the viability of the BPS Skyrme model as the leading contribution to an effective theory for the properties of nuclear matter. Their specific choice of the potential was motivated by the two requirements to avoid the compact nature of the solitons, and to have the standard pion tail in the full model with the quadratic Skyrme term included.
It is this point where we slightly disagree, because we think that these two requirements are mutually incompatible (this issue is, however, completely irrelevant for the remaining results of that paper). Their potential has, in fact, two vacua and behaves quadratically about one vacuum, but sextic around the other.  Further, their solitons approach the sextic vacuum at large distances. This implies that the pure BPS model solutions are not compact, but in the full model there is no pion tail (the pion tail would be induced by the quadratic vacuum which is, however, assumed at the center of the soliton, and not in the asymptotic large distance region).
In any case, further investigations of the semiclassical quantization with the inclusion of higher excitations of nuclei and electrostatic effects, among others, would be very interesting.
\item Motivation from QCD: it would be very interesting to see whether the rather good phenomenological properties of the model can be justified in a more rigorous manner from the fundamental theory of strong interactions, i.e., QDC. In this context we want to emphasize that the two term of the BPS Skyrme model are rather specific. The sextic term is essentially topological in nature and, therefore, naturally related to collective excitations of degrees of freedom of the underlying microscopic theory. The potential, on the other hand, provides the chiral symmetry breaking and might therefore be related to collective degrees of freedom of the quarks, like the quark condensate. These issues certainly require further investigation.
\end{itemize}
In any case, we think that we have found and solved
an important submodel in the
space of generalized Skyrme models, 
which is singled out both by its
capacity to reproduce qualitative properties of the liquid drop
model of nuclei and by its unique 
mathematical structure.
The model directly relates the nuclear mass to the topological charge, and it
naturally provides both a finite size for the nuclei and the liquid drop
behaviour, which probably is not easy to get from an effective field
theory. 
Finally, our exact solutions might provide a calibration for the 
demanding numerical computations in physical applications of more 
general Skyrme models.

\section*{Acknowledgements}

C.A. and J.S.-G. thank the Ministry of Science and Investigation, Spain 
(grant FPA2008-01177), and
the Xunta de Galicia (grant INCITE09.296.035PR and
Conselleria de Educacion) for financial support.
A.W. acknowledges support from the
Ministry of Science and Higher Education of Poland grant N N202
126735 (2008-2010).

 \end{document}